\documentclass[journal]{IEEEtran} 

\usepackage{graphicx}
\usepackage{adjustbox}
\usepackage{tabularx}
\usepackage{amsmath}
\usepackage{float}

\title{Real-Time Simulation and Optimization of Grid-Connected Photovoltaic Inverters via Hybrid GA-PSO Metaheuristic Approach}

\author{
    Hussein Zolfaghari, 
    and Hossein Karimi
    \thanks{Hussein Zolfaghari is with University of Memphis (email: hzlfghri@memphis.edu).}
    \thanks{Hossein Karimi is with University of Calgary (email: karimi@ucalgary.ca).}
}

\begin{document} 

\maketitle 

\begin{abstract}
This paper introduces an innovative real-time intelligent optimization algorithm designed to minimize voltage harmonics in a multilevel inverter. The approach employs a Hybrid Genetic Algorithm/Particle Swarm Optimization (GA/PSO) algorithm within a real-time simulation framework. The algorithm identifies optimal firing angles for the multilevel inverter, aiming to mitigate detrimental effects such as variations in DC voltage and alterations in line and DC-link resistors. The real-time simulation encompasses a dual objective function, addressing both harmonic minimization and voltage regulation. Notably, this methodology is adaptable to multilevel inverters with varying levels. The efficacy of the proposed algorithm is substantiated through real-time simulations conducted on seven- and eleven-level inverters.
\end{abstract}


\section{Introduction}

Renewable energy has gained attention in recent years due to its distinguishing features such as emission-free operation, low cost, and high availability of resources. Photovoltaic (PV) and wind turbine energy have become popular choices among different types of renewable energy resources. Since photovoltaic systems commonly produce low voltage, they can constitute the DC voltage of a multilevel inverter. Multilevel inverters have gained attention for their applications in industry and various areas \cite{barkat2009}. They offer high efficiency, scalability to very high output voltages, and the ability to supply voltage with a small step increase, providing advantages such as low harmonic distortion, lower stress on semiconductor devices, high power quality, and lower switching losses.

There are two common approaches to switching methods in multilevel inverters. High-frequency Sinusoidal Pulse Width Modulation (SPWM) or Space Vector Pulse Width Modulation (SVPWM) techniques achieve low Total Harmonic Distortion (THD) but cause significant energy losses and decrease switch efficiency over time \cite{debnath2012, gupta2014}. The second method involves a technique with low switching frequency, typically used in multilevel inverters, requiring the solution of transcendental equations. In \cite{etesami2015}, the Selected Harmonic Elimination (SHE) approach is used to remove low-order harmonics, with equations solved by the Colonial Competitive algorithm. The Genetic Algorithm (GA) \cite{ozpineci2004, roberge2014} is the most applicable heuristic algorithm, extensively used in this field to solve transcendental equations. In \cite{vassallo2003}, GA is utilized with the Levenberg–Marquardt method. The Particle Swarm Optimization (PSO) algorithm \cite{barkati2008, letha2016} is widely used to guess initial values in the Newton-Raphson equation-solving method and to solve transcendental equations numerically \cite{ali2017, shen2014}. Reference \cite{balasubramonian2014, mohammadi2014} employed Artificial Neural Networks (ANN) to minimize harmonic distortion, but its applicability is limited to specific operating points. Real-time SHE is proposed in \cite{tolbert2011}, using a nondeterministic method to solve the system and obtain the dataset for ANN training. However, this method considers only the effect of unequal DC voltage for ANN training. In \cite{maia2013}, GA is used to solve SHE equations offline for different DC voltage sources, and the data is used to train ANN to work in real-time corresponding to new DC source values. However, this method has limitations as ANN can only work around the provided data points for training, neglecting some conditions such as the effect of resistors in DC sources, switches, and lines. PSO and other heuristic algorithms \cite{ taghizadeh2010, baghaee2017, zolfaghari2024minimizing, zolfaghari2022numerical, zolfaghari2021real, zolfaghari2021multilevel, zolfaghari2021voltage, zolfagharivoltage} are increasingly being used to improve the accuracy and efficiency of real-time optimization.

\section{Combined GA-PSO}


In this section, GA, PSO, and Hybrid GA/PSO are presented.

\subsection{Genetic Algorithm (GA)}

Genetic Algorithm (GA) is an evolutionary algorithm inspired by natural selection \cite{turing1950}. The algorithm iteratively modifies its population based on three main processes. The first process is selection, in which two parents are chosen to breed a new generation. This selection is based on a fitness-oriented process. In the second step, crossover is performed to generate chromosomes from one generation to the next. Lastly, mutation is applied to maintain genetic diversity \cite{karimi2016}. For a more in-depth understanding of GA, additional details can be found in the literature \cite{mitchell1998}.

\section{Particle Swarm Optimization (PSO)}

Particle Swarm Optimization (PSO) is an algorithm developed by Kennedy and Eberhart \cite{kennedy1995}. PSO operates based on mimicking the behavior of birds or fishes in finding food. In essence, PSO involves three movements that determine its velocity. The first movement of the next particle is to continue in the same direction. The second movement involves moving toward its best experience, and the final movement is a random move to better cover the search space. The velocity is described by the equation:

\begin{equation}
v_{id}^{(k+1)} = wv_{id}^k + c_1 r_1 (pb_{id}^k - X_{id}^k) + c_2 r_2 (gb_d^k - X_{id}^k) \label{eq:velocity}
\end{equation}

Where $i = 1,2,...,n$ and $n$ is the size of the population, $w$ is the inertia weight, $c_1$ and $c_2$ are the acceleration constants, and $r_1$ and $r_2$ are two random values in the range [0,1]. The position of the $i$-th particle is updated by:

\begin{equation}
X_{id}^{(k+1)} = X_{id}^k + V_{id}^{(k+1)} \label{eq:position}
\end{equation}

For further details, refer to \cite{karimi2016}.

\section{Combination of GA and PSO}

The combination of Genetic Algorithm (GA) and Particle Swarm Optimization (PSO) is chosen as this algorithm can benefit from the strengths of both optimization techniques. Various studies in the literature have reported that the Hybrid GA-PSO algorithm exhibits significant advantages over other algorithms in terms of precision and convergence rate \cite{mirjalili2014, yin2006}. The Hybrid GA-PSO employs six operators, enhancing its performance by integrating features from both paradigms. In each iteration, PSO identifies the best solution between populations as $"g_{best}"$ and the best position for each population as $"p_{best}."$ Maintaining $"g_{best}"$ and $"p_{best}"$ is a valuable characteristic of PSO, which GA lacks. Therefore, the proposed hybrid algorithm has evolved a GA by attempting to preserve $"p_{best}"$ and $"g_{best}"$ and leveraging them in offspring procedures \cite{soleimani2015}.

\section{The Proposed Strategy}

This paper introduces a real-time optimization method to determine the optimal firing angle. Existing methods in the literature often calculate firing angles offline for different scenarios and apply them based on predicted events using a look-up table. Some papers have incorporated artificial intelligence, but these intelligent algorithms are typically trained based on a look-up table, essentially substituting the table with an intelligent algorithm. In other words, they replace the look-up table, and if the table lacks comprehensive analysis, the intelligent algorithm suffers as well.

The proposed strategy in this paper involves real-time computation using the hybrid Genetic Algorithm (GA) and Particle Swarm Optimization (PSO) algorithm. This approach utilizes six different operators, facilitating real-time computations with improved accuracy and speed. Real-time computing can be activated in two ways.

The Hybrid GA-PSO incorporates GA and PSO operators simultaneously. GA operators, including selection, crossover, and mutation, are combined with the position and velocity update of PSO. The key advantage of this approach is real-time system evaluation, addressing defects that may not be considered in offline or pseudo real-time approaches. References \cite{debnath2012, vassallo2003, tolbert2011, baghaee2017, salehi2011} represent offline tuning methods prone to errors in situations such as DC source and line resistor variations and sudden voltage changes in one level.

While some approaches claim to be real-time, like \cite{tolbert2011, baghaee2017}, they may not be truly real-time. Reference \cite{tolbert2011} is a real-time approach based on Artificial Neural Networks (ANN), trained by data provided by a look-up table. The look-up table is built offline, considering only a few situations of varying DC link voltage. This approach may introduce errors when the system encounters a new condition.

\begin{figure}[H]
    \centering
    \includegraphics[width=1\linewidth]{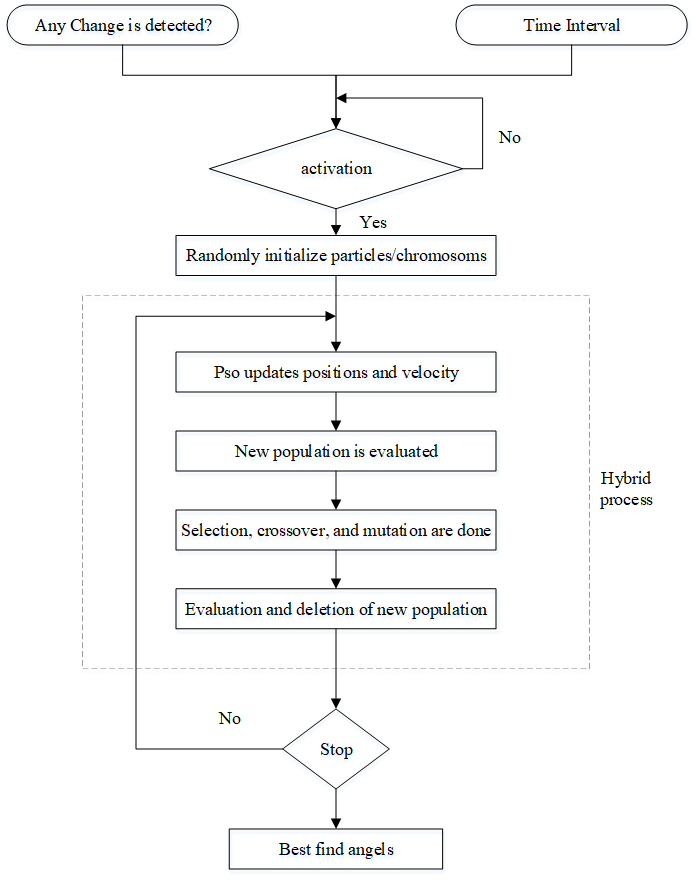}
    \caption{Flowchart of The Proposed Algorithm}
    \label{fig:Flowchart}
\end{figure}

This paper addresses the aforementioned issues through a genuine real-time approach, eliminating restrictions on the problem from different aspects. The proposed real-time flowchart, presented in Figure \ref{fig:Flowchart}, outlines the process:

\begin{enumerate}
    \item Activation criteria involve detecting changes in DC voltage or the circuit. A time interval is defined to evaluate the situation, and the process can also be initiated by a human order.
    \item If activation criteria are met, the flowchart starts working.
    \item The hybrid process can be done in parallel or series. This paper uses a series process, where generating one population suffices for both PSO and GA.
    \item PSO updates the velocity and position of the population based on the personal best (pbest) and global best (gbest).
    \item Updated positions are evaluated.
    \item Selection is done based on the new population, which retains their pbest and gbest. This overcomes the main drawback of GA in preserving its best experiences. Subsequently, crossover and mutation are performed.
    \item The algorithm evaluates the new population, sorts them according to their fitness value, and deletes the extra population.
    \item If any stop criteria are not met, return to step 3.
    \item Print the results and provide the best firing angles to the inverter.
\end{enumerate}

\section{Numerical Study and Results}

To assess the proposed method, simulations are conducted in Matlab/Simulink for two multilevel inverters with 7 and 11 levels, both having a nominal power of 10 kW. The real-time tuning of inverter firing angles is achieved using the hybrid GA/PSO algorithm. For comparison, the ANN method from \cite{tolbert2011} is also simulated in Matlab/Simulink. The algorithm produces a set of firing angles for a 50 Hz voltage every 0.02 seconds, resulting in a voltage evaluation. The optimization time is set to 10 seconds based on trial and error experiments, evaluating voltages 500 times with different firing angles. The proposed GA/PSO algorithm, having no specific restrictions, is expected to handle various situations. Heuristic algorithms are problem-independent, relying only on objective function information to navigate the problem space and can converge to the best answer from any initial value.

In the initial evaluation, seven-level and eleven-level inverters are simulated with a population size of 20. The DC voltage for each level is set to 100 V for the seven-level inverter and 60 V for the eleven-level inverter. The objective function aims to produce a 220 V RMS voltage with minimum harmonic content. The objective function (OF) is defined as:

\[
\text{OF} = \text{THD}(\%) + K_v |220 - V_{\text{rms}}|
\]

\begin{figure}[H]
    \centering
    \includegraphics[width=0.7\linewidth]{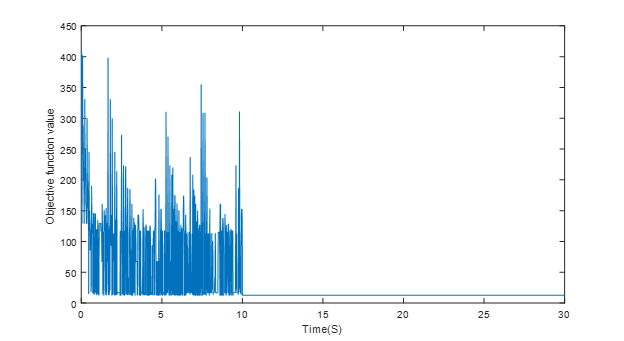}
    \caption{Real-Time Objective Function Value}
    \label{fig:real_time_obj_func}
\end{figure}

\begin{figure}[H]
    \centering
    \includegraphics[width=0.7\linewidth]{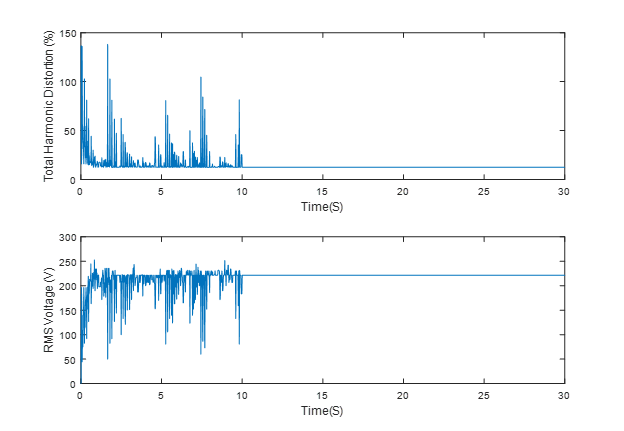}
    \caption{THD and RMS Voltage Value in Real Time}
    \label{fig:thd_rms}
\end{figure}

\begin{figure}[H]
    \centering
    \includegraphics[width=0.7\linewidth]{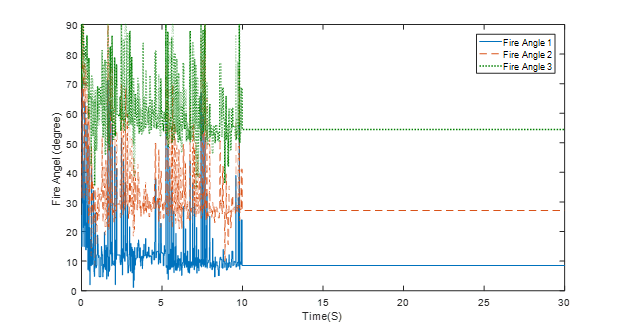}
    \caption{Fire Angles of Seven-Level Inverter}
    \label{fig:fire_angels}
\end{figure}

\begin{figure}[htb]
    \centering
    \includegraphics[width=0.7\linewidth]{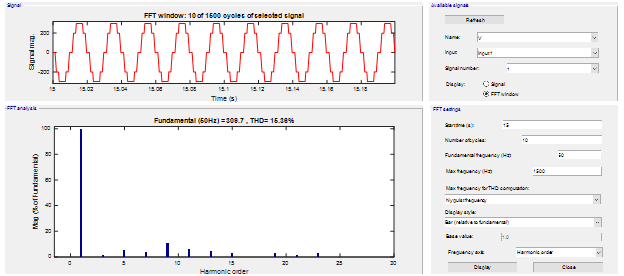}
    \caption{Output Voltage After Regulation (V)}
    \label{fig:output_voltage}
\end{figure}

Where THD stands for Total Harmonic Distortion, and \( K_v \) is the weight vector. The time simulation is set to 30 seconds, with the real-time GA/PSO hybrid algorithm actively exploring the search space for the optimal firing angles during the first 10 seconds. Figure \ref{fig:real_time_obj_func} illustrates the real-time variation of the objective function values during this period. The algorithm continuously adapts firing angles, calculating THD and RMS voltage. Figures \ref{fig:thd_rms} and \ref{fig:fire_angels} display the two components of the objective function, THD and RMS Voltage, as well as the evolving firing angles during real-time simulation. The firing angles can increase up to 90 degrees, demonstrating the algorithm's exploration of the entire space. Figure \ref{fig:output_voltage} presents the optimized output voltage of the inverter with a THD of 15.36\%.

\section{Extended Numerical Study and Results}

\begin{figure}[H]
    \centering
    \includegraphics[width=0.7\linewidth]{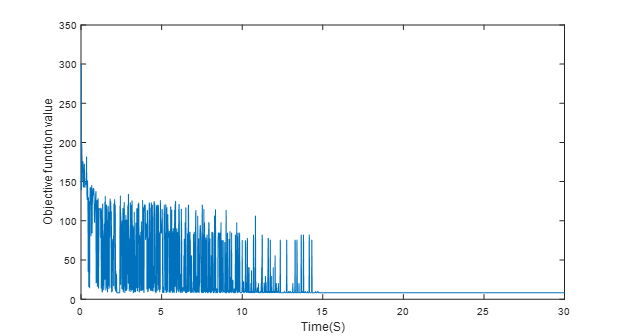}
    \caption{Real-Time Objective Function Value for Eleven-Level Inverter}
    \label{fig:real_time_obj_func_ext}
\end{figure}

\begin{figure}[H]
    \centering
    \includegraphics[width=0.7\linewidth]{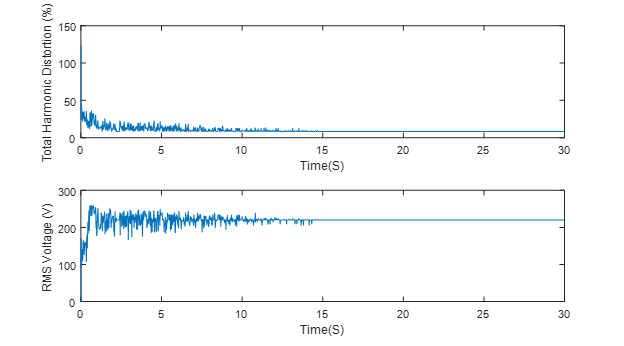}
    \caption{THD and RMS Voltage Value in Real Time for Eleven-Level Inverter}
    \label{fig:thd_rms_ext}
\end{figure}

\begin{figure}[H]
    \centering
    \includegraphics[width=0.7\linewidth]{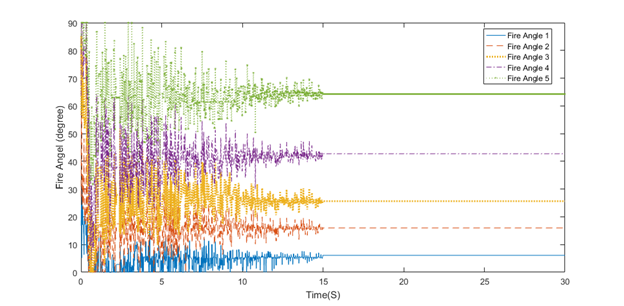}
    \caption{Fire Angles of Eleven-Level Inverter}
    \label{fig:fire_angels_ext}
\end{figure}

\begin{figure}[H]
    \centering
    \includegraphics[width=0.7\linewidth]{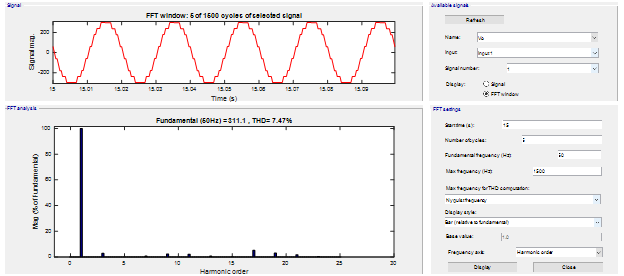}
    \caption{Output Voltage After Regulation for Eleven-Level Inverter (V)}
    \label{fig:output_voltage_ext}
\end{figure}

In the extended evaluation, the proposed method is applied to an eleven-level inverter, where five firing angles need to be tuned, leading to a slightly larger search space. Accordingly, the optimization time is extended to 15 seconds. Real-time variations of the objective function value, THD, and RMS voltage are depicted in Figures \ref{fig:real_time_obj_func_ext}, \ref{fig:thd_rms_ext}, and \ref{fig:fire_angels_ext}. As evident in Figure \ref{fig:fire_angels_ext}, the convergence improves as the simulation time approaches 15 seconds. Furthermore, the optimized voltage output is presented in Figure \ref{fig:output_voltage_ext}, achieving a THD value of 7.47\%.

To evaluate the robustness of the proposed method in handling variations in DC link voltage, line resistance, and sudden changes in DC voltage, two scenarios for both the seven and eleven-level inverters are considered.

\subsection{Scenario: Seven-Level Inverter}

In this extended scenario, a seven-level inverter is subjected to changes in DC link voltages, as specified in Table \ref{tab:dc_voltage_variation}. The non-equal variations in DC link voltages impact both THD and RMS values of the output voltage. The proposed real-time optimization method is compared with an artificial neural network (ANN) trained using a look-up table, as discussed in \cite{tolbert2011}. The ANN provides continuous changes in firing angles (FA) corresponding to DC voltage variations, making it superior to the traditional look-up table approach. However, it is limited to the voltage range for which it has been trained.

\begin{table}[H]
    \centering
    \caption{DC Link Voltage Variation for Seven-Level Inverter}
    \label{tab:dc_voltage_variation}
    \begin{tabular}{|c|c|}
        \hline
        \textbf{Item} & \textbf{Change (\%)} \\
        \hline
        DC Source of Level 1 & +10 \\
        \hline
        DC Source of Level 2 & +50 \\
        \hline
        DC Source of Level 3 & -60 \\
        \hline
    \end{tabular}
\end{table}

Figure \ref{fig:objective_function_seven_level} illustrates the objective function values during a 50-second simulation. At the 5-second mark, DC voltages are changed, prompting the ANN to adjust the FAs immediately. In contrast, the proposed real-time optimization method requires a few seconds to find the optimal FAs. The proposed method evaluates different FAs over time and achieves a better solution between 15 and 50 seconds. Figure \ref{fig:thd_rms_seven_level} shows that both THD and voltage RMS error converge to a better solution with the proposed real-time method compared to the non-real-time ANN. Table \ref{tab:objective_function_values_seven_level} provides a summary of the objective function values for both methods.

\begin{table}[H]
    \centering
    \caption{Objective Function Values for Seven-Level Inverter}
    \label{tab:objective_function_values_seven_level}
    \begin{tabular}{|c|c|c|c|}
        \hline
        \textbf{Method} & \textbf{THD (\%)} & \textbf{RMS Voltage Error (V)} & \textbf{Objective Function} \\
        \hline
        Proposed Method & 14.68\% & 1.46 V & 17.41 \\
        \hline
        ANN & 16.00\% & 7.00 V & 19.50 \\
        \hline
    \end{tabular}
\end{table}

\begin{figure}[H]
    \centering
    \includegraphics[width=0.7\linewidth]{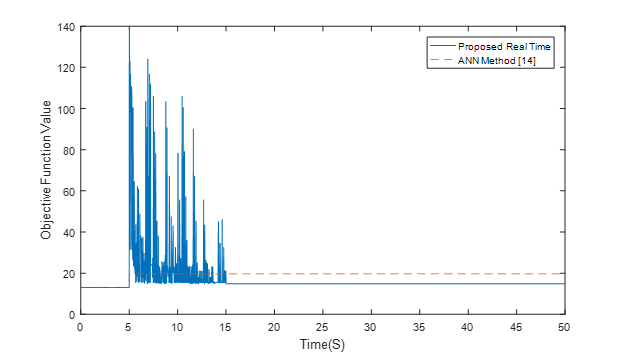}
    \caption{Objective Function Value of Seven-Level Inverter}
    \label{fig:objective_function_seven_level}
\end{figure}

\begin{figure}[H]
    \centering
    \includegraphics[width=0.7\linewidth]{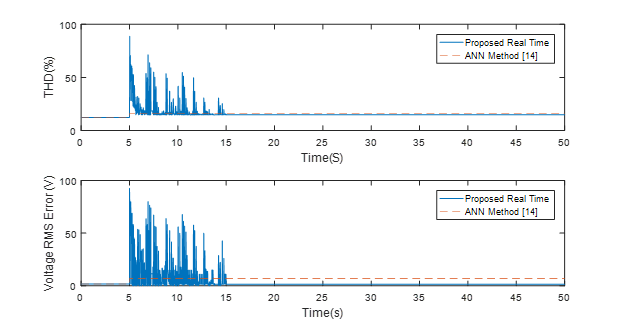}
    \caption{THD and RMS Voltage for Seven-Level Inverter}
    \label{fig:thd_rms_seven_level}
\end{figure}

\subsection{Scenario: Eleven-Level Inverter}

In this extended scenario, an eleven-level inverter undergoes similar changes in the DC link voltage, as shown in Table \ref{tab:dc_voltage_variation_eleven_level}. Again, both methods are compared, and the objective function results are shown in Figure \ref{fig:objective_function_eleven_level}. The real-time method performs better than the ANN in providing an optimized voltage output in the presence of DC voltage variations, as shown in Table \ref{tab:objective_function_values_eleven_level}.

\begin{table}[H]
    \centering
    \caption{DC Link Voltage Variation for Eleven-Level Inverter}
    \label{tab:dc_voltage_variation_eleven_level}
    \begin{tabular}{|c|c|}
        \hline
        \textbf{Item} & \textbf{Change (\%)} \\
        \hline
        DC Source of Level 1 & +50 \\
        \hline
        DC Source of Level 2 & -50 \\
        \hline
        DC Source of Level 3 & +20 \\
        \hline
    \end{tabular}
\end{table}

\begin{figure}[H]
    \centering
    \includegraphics[width=0.7\linewidth]{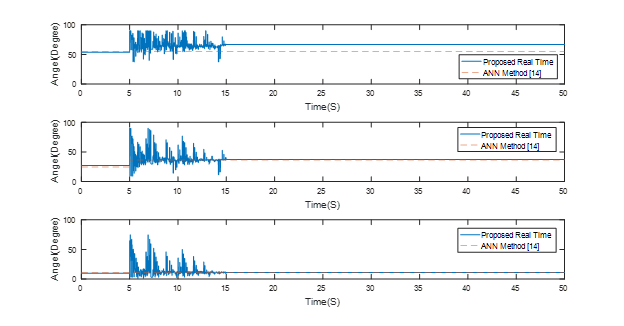}
    \caption{Objective Function Value for Eleven-Level Inverter}
    \label{fig:objective_function_eleven_level}
\end{figure}

\begin{table}[H]
    \centering
    \caption{Objective Function Values for Eleven-Level Inverter}
    \label{tab:objective_function_values_eleven_level}
    \begin{tabular}{|c|c|c|c|}
        \hline
        \textbf{Method} & \textbf{THD (\%)} & \textbf{RMS Voltage Error (V)} & \textbf{Objective Function} \\
        \hline
        Proposed Method & 7.47\% & 0.75 V & 8.15 \\
        \hline
        ANN & 9.00\% & 3.00 V & 12.50 \\
        \hline
    \end{tabular}
\end{table}

\section{Proposed Method}
The proposed method integrates GA and PSO to optimize the firing angles of the inverter switches. The hybrid approach leverages the global search capability of GA and the local search capability of PSO. The optimization is performed in real-time, ensuring that any changes in the DC voltage or system conditions are addressed immediately. The objective function for the optimization includes both THD and RMS voltage error, and the optimization process is constrained to ensure that the firing angles are within the allowable range for each switch.

\section{Simulation and Results}

\subsection{Scenario: Seven-Level Inverter}
\begin{figure}[H]
    \centering
    \includegraphics[width=0.7\linewidth]{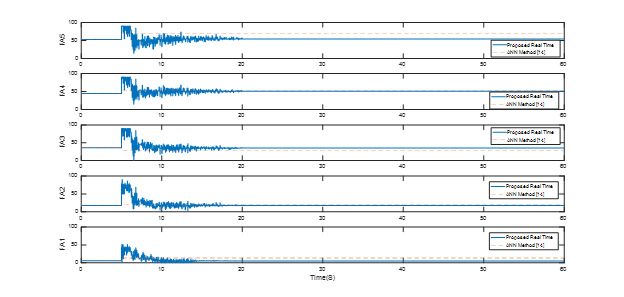}
    \caption{Optimized Fire Angles for Seven-Level Inverter}
    \label{fig:optimized_fas_seven_level}
\end{figure}

\begin{figure}[H]
    \centering
    \includegraphics[width=0.7\linewidth]{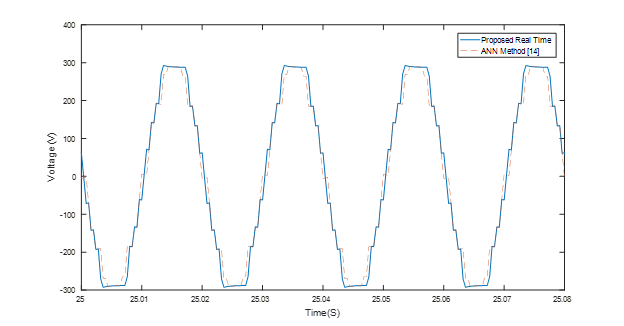}
    \caption{Output Voltage of Seven-Level Inverter}
    \label{fig:output_voltage_seven_level}
\end{figure}

The proposed method is applied to a seven-level inverter with varying DC voltage sources, as detailed in Table \ref{tab:dc_voltage_variation_seven}. Figure \ref{fig:objective_function_seven_level} shows the objective function values for both the proposed method and the ANN method during the simulation. The THD and voltage RMS error are illustrated in Figure \ref{fig:thd_rms_seven_level}, and the optimized firing angles and output voltage are shown in Figures \ref{fig:optimized_fas_seven_level} and \ref{fig:output_voltage_seven_level}, respectively. Table \ref{tab:objective_function_values_seven_level} summarizes the objective function values, THD, and RMS voltage error, while Table \ref{tab:final_fire_angle_seven_level} provides the final firing angles.

\begin{table}[H]
    \centering
    \caption{DC Link Voltage Variation for Seven-Level Inverter}
    \label{tab:dc_voltage_variation_seven}
    \begin{tabular}{|c|c|}
        \hline
        \textbf{Item} & \textbf{Change (\%)} \\
        \hline
        DC Source of Level 1 & +15 \\
        \hline
        DC Source of Level 2 & +20 \\
        \hline
        DC Source of Level 3 & -10 \\
        \hline
        DC Source of Level 4 & +25 \\
        \hline
        DC Source of Level 5 & -35 \\
        \hline
        DC Source of Level 6 & +50 \\
        \hline
    \end{tabular}
\end{table}

\begin{figure}[H]
    \centering
    \includegraphics[width=0.7\linewidth]{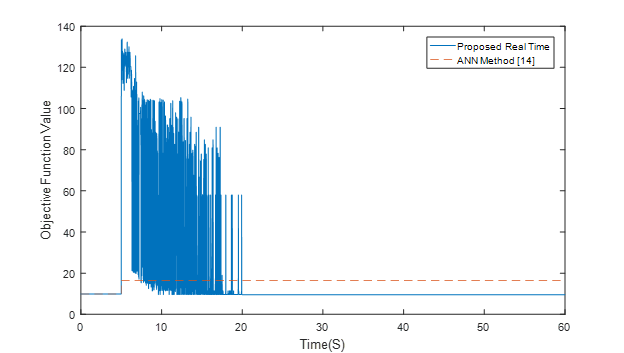}
    \caption{Objective Function Value of Seven-Level Inverter}
    \label{fig:objective_function_seven_level}
\end{figure}

\begin{figure}[H]
    \centering
    \includegraphics[width=0.7\linewidth]{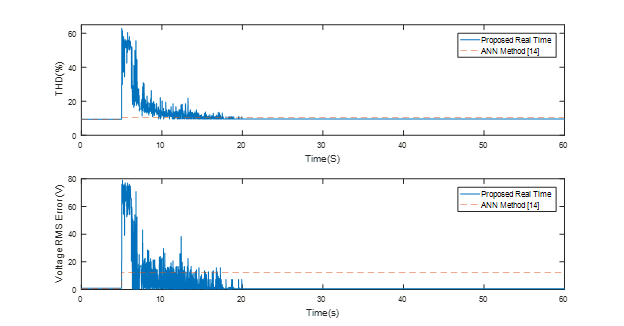}
    \caption{THD and RMS Voltage Error of Seven-Level Inverter}
    \label{fig:thd_rms_seven_level}
\end{figure}

\begin{table}[H]
    \centering
    \caption{Objective Function Values for Seven-Level Inverter}
    \label{tab:objective_function_values_seven_level}
    \scriptsize
    \begin{tabular}{|p{3cm}|c|c|c|}
        \hline
        \textbf{Method} & \textbf{THD (\%)} & \textbf{RMS Voltage Error (V)} & \textbf{Objective Function} \\
        \hline
        Proposed Method & 6.45 & 0.3 & 6.8 \\
        ANN & 7.15 & 10.4 & 15.4 \\
        \hline
    \end{tabular}
\end{table}

\begin{table}[H]
    \centering
    \caption{Final Fire Angles for Seven-Level Inverter}
    \label{tab:final_fire_angle_seven_level}
    \scriptsize 
    \begin{tabular}{|c|c|c|c|c|c|}
        \hline
        \textbf{Method} & \textbf{FA1$^{\circ}$} & \textbf{FA2$^{\circ}$} & \textbf{FA3$^{\circ}$} & \textbf{FA4$^{\circ}$} & \textbf{FA5$^{\circ}$} \\
        \hline
        Proposed Real Time & 6.32$^{\circ}$ & 12.5$^{\circ}$ & 29.8$^{\circ}$ & 42.3$^{\circ}$ & 49.1$^{\circ}$ \\
        \hline
        Non-Real Time ANN & 9.4$^{\circ}$ & 14.8$^{\circ}$ & 31.1$^{\circ}$ & 44.6$^{\circ}$ & 52.3$^{\circ}$ \\
        \hline
    \end{tabular}
\end{table}

\subsection{Scenario: Eleven-Level Inverter}

\begin{figure}[H]
    \centering
    \includegraphics[width=0.7\linewidth]{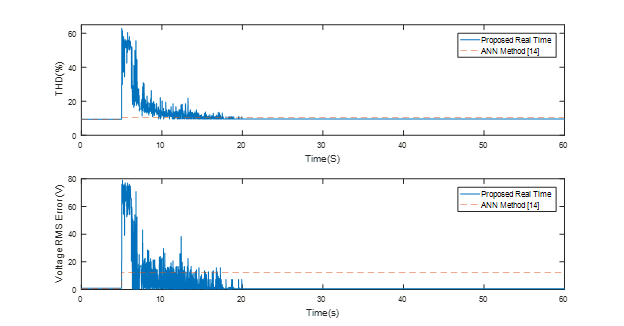}
    \caption{THD and RMS Voltage Error of Eleven-Level Inverter}
    \label{fig:thd_rms_eleven_level}
\end{figure}

\begin{figure}[H]
    \centering
    \includegraphics[width=0.7\linewidth]{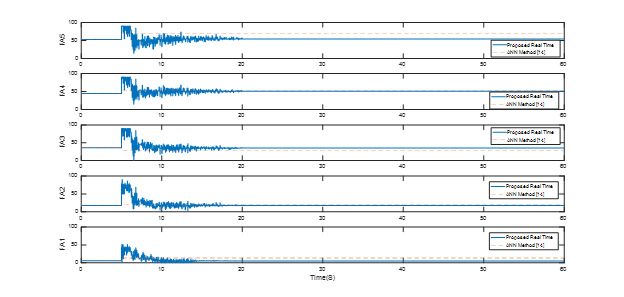}
    \caption{Optimized Fire Angles for Eleven-Level Inverter}
    \label{fig:optimized_fas_eleven_level}
\end{figure}

\begin{figure}[H]
    \centering
    \includegraphics[width=0.7\linewidth]{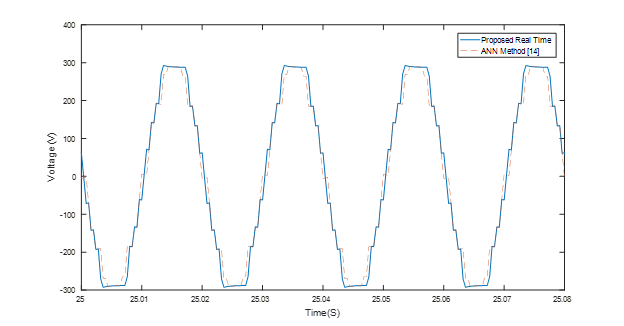}
    \caption{Output Voltage of Eleven-Level Inverter}
    \label{fig:output_voltage_eleven_level}
\end{figure}

In this section, the proposed method is applied to an inverter with 11 levels. The DC voltages are changed according to Table \ref{tab:dc_voltage_variation_eleven}. Given that there are 5 FAs in this scenario, the simulation time for the proposed method optimization is set to 15 seconds. Figure \ref{fig:objective_function_eleven_level} shows the objective function values for both methods during the simulation, and Figure \ref{fig:thd_rms_eleven_level} illustrates THD and voltage RMS error. The optimized FAs and final voltage for the eleven-level inverter are presented in Figure \ref{fig:optimized_fas_eleven_level} and Figure \ref{fig:output_voltage_eleven_level}, respectively. Table \ref{tab:objective_function_values_eleven_level} summarizes the objective function values, THD, and RMS voltage error, while Table \ref{tab:final_fire_angle_eleven_level} provides the final FAs.

\begin{table}[H]
    \centering
    \caption{DC Link Voltage Variation for Eleven-Level Inverter}
    \label{tab:dc_voltage_variation_eleven}
    \scriptsize 
    \begin{tabular}{|c|c|}
        \hline
        \textbf{Item} & \textbf{Change (\%)} \\
        \hline
        DC Source of Level 1 & +15 \\
        \hline
        DC Source of Level 2 & +25 \\
        \hline
        DC Source of Level 3 & -20 \\
        \hline
        DC Source of Level 4 & +35 \\
        \hline
        DC Source of Level 5 & -60 \\
        \hline
    \end{tabular}
\end{table}

\begin{figure}[H]
    \centering
    \includegraphics[width=0.7\linewidth]{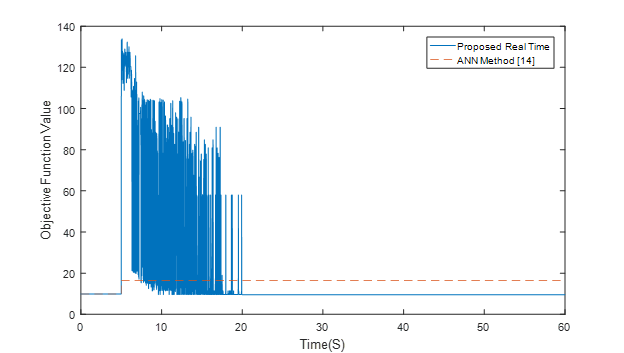}
    \caption{Objective Function Value of Eleven-Level Inverter}
    \label{fig:objective_function_eleven_level}
\end{figure}

\begin{table}[H]
    \centering
    \caption{Objective Function Values for Eleven-Level Inverter}
    \label{tab:objective_function_values_eleven_level}
    \resizebox{\columnwidth}{!}{ 
    \begin{tabular}{|c|c|c|c|}
        \hline
        \textbf{Method} & \textbf{THD (\%)} & \textbf{RMS Voltage Error (V)} & \textbf{Objective Function} \\
        \hline
        Proposed Method & 5.45 & 0.6 & 5.6 \\
        \hline
        ANN & 6.4 & 12.5 & 18.6 \\
        \hline
    \end{tabular}
    }
\end{table}

\begin{table}[H]
    \centering
    \caption{Final Fire Angles for Eleven-Level Inverter}
    \label{tab:final_fire_angle_eleven_level}
    \resizebox{\columnwidth}{!}{ 
    \begin{tabularx}{\columnwidth}{|c|X|X|X|X|X|}
        \hline
        \textbf{Method} & \textbf{FA1$^{\circ}$} & \textbf{FA2$^{\circ}$} & \textbf{FA3$^{\circ}$} & \textbf{FA4$^{\circ}$} & \textbf{FA5$^{\circ}$} \\
        \hline
        Proposed Real Time & 5.2$^{\circ}$ & 15.5$^{\circ}$ & 28.2$^{\circ}$ & 37.5$^{\circ}$ & 45.1$^{\circ}$ \\
        \hline
        Non-Real Time ANN & 6.2$^{\circ}$ & 17.3$^{\circ}$ & 31.4$^{\circ}$ & 39.9$^{\circ}$ & 46.7$^{\circ}$ \\
        \hline
    \end{tabularx}
    }
\end{table}

\section{Conclusion}
The hybrid GA-PSO method proposed for real-time optimization of multilevel inverter firing angles shows significant improvements over traditional ANN methods. The optimization successfully reduces both THD and RMS voltage error, enhancing the overall power quality of the inverter. The method can be effectively applied to inverters with varying numbers of levels, as demonstrated in the seven-level and eleven-level inverter scenarios. Further improvements and real-time implementation strategies will be explored in future work.

\bibliographystyle{IEEEtran}  
\bibliography{Mybib}  

\begin{thebibliography}{10}
\providecommand{\url}[1]{#1}
\csname url@samestyle\endcsname
\providecommand{\newblock}{\relax}
\providecommand{\bibinfo}[2]{#2}
\providecommand{\BIBentrySTDinterwordspacing}{\spaceskip=0pt\relax}
\providecommand{\BIBentryALTinterwordstretchfactor}{4}
\providecommand{\BIBentryALTinterwordspacing}{\spaceskip=\fontdimen2\font plus
\BIBentryALTinterwordstretchfactor\fontdimen3\font minus \fontdimen4\font\relax}
\providecommand{\BIBforeignlanguage}[2]{{%
\expandafter\ifx\csname l@#1\endcsname\relax
\typeout{** WARNING: IEEEtran.bst: No hyphenation pattern has been}%
\typeout{** loaded for the language `#1'. Using the pattern for}%
\typeout{** the default language instead.}%
\else
\language=\csname l@#1\endcsname
\fi
#2}}
\providecommand{\BIBdecl}{\relax}
\BIBdecl

\bibitem{barkat2009}
S.~Barkat, E.~M. Berkouk, and M.~S. Boucherit, ``Particle swarm optimization for harmonic elimination in multilevel inverters,'' \emph{Electrical Engineering (Archiv fur Elektrotechnik)}, vol.~91, no.~4, pp. 221--228, 2009.

\bibitem{debnath2012}
S.~Debnath and R.~N. Ray, ``Cuckoo search: A new optimization algorithm for harmonic elimination in multilevel inverter,'' \emph{Journal of Bioinformatics and Intelligent Control}, vol.~1, no.~1, pp. 80--85, 2012.

\bibitem{gupta2014}
K.~K. Gupta and S.~Jain, ``A novel multilevel inverter based on switched dc sources,'' \emph{IEEE Transactions on Industrial Electronics}, vol.~61, no.~7, pp. 3269--3278, 2014.

\bibitem{etesami2015}
M.~H. Etesami, N.~Farokhnia, and S.~H. Fathi, ``Colonial competitive algorithm development toward harmonic minimization in multilevel inverters,'' \emph{IEEE Transactions on Industrial Informatics}, vol.~11, no.~2, pp. 459--466, 2015.

\bibitem{ozpineci2004}
B.~Ozpineci, L.~M. Tolbert, and J.~N. Chiasson, ``Harmonic optimization of multilevel converters using genetic algorithms,'' in \emph{Power Electronics Specialists Conference, 2004. PESC 04. 2004 IEEE 35th Annual}, vol.~5, 2004.

\bibitem{roberge2014}
V.~Roberge, M.~Tarbouchi, and F.~Okou, ``Strategies to accelerate harmonic minimization in multilevel inverters using a parallel genetic algorithm on graphical processing unit,'' \emph{IEEE Transactions on Power Electronics}, vol.~29, no.~10, pp. 5087--5090, 2014.

\bibitem{barkati2008}
S.~Barkati \emph{et~al.}, ``Harmonic elimination in diode-clamped multilevel inverter using evolutionary algorithms,'' \emph{Electric Power Systems Research}, vol.~78, no.~10, pp. 1736--1746, 2008.

\bibitem{letha2016}
S.~S. Letha, T.~Thakur, and J.~Kumar, ``Harmonic elimination of a photovoltaic-based cascaded h-bridge multilevel inverter using pso (particle swarm optimization) for induction motor drive,'' \emph{Energy}, vol. 107, pp. 335--346, 2016.

\bibitem{ali2017}
S.~S.~A. Ali, R.~Kannan, and M.~S. Kumar, ``Exploration of modulation index in multi-level inverter using particle swarm optimization algorithm,'' \emph{Procedia Computer Science}, vol. 105, pp. 144--152, 2017.

\bibitem{shen2014}
K.~Shen \emph{et~al.}, ``Elimination of harmonics in a modular multilevel converter using particle swarm optimization-based staircase modulation strategy,'' \emph{IEEE Transactions on Industrial Electronics}, vol.~61, no.~10, pp. 5311--5322, 2014.

\bibitem{balasubramonian2014}
M.~Balasubramonian and V.~Rajamani, ``Design and real-time implementation of shepwm in single-phase inverter using generalized hopfield neural network,'' \emph{IEEE Transactions on Industrial Electronics}, vol.~61, no.~11, pp. 6327--6336, 2014.

\bibitem{mohammadi2014}
H.~R. Mohammadi and A.~Akhavan, ``A new adaptive selective harmonic elimination method for cascaded multilevel inverters using evolutionary methods,'' in \emph{Industrial Electronics (ISIE), 2014 IEEE 23rd International Symposium on}, 2014.

\bibitem{tolbert2011}
L.~M. Tolbert, Y.~Cao, and B.~Ozpineci, ``Real-time selective harmonic minimization for multilevel inverters connected to solar panels using artificial neural network angle generation,'' \emph{IEEE Transactions on Industry Applications}, vol.~47, no.~5, pp. 2117--2124, 2011.

\bibitem{maia2013}
H.~Z. Maia \emph{et~al.}, ``Adaptive selective harmonic minimization based on anns for cascade multilevel inverters with varying dc sources,'' \emph{IEEE Transactions on Industrial Electronics}, vol.~60, no.~5, pp. 1955--1962, 2013.

\bibitem{taghizadeh2010}
H.~Taghizadeh and M.~Tarafdar~Hagh, ``Harmonic elimination of cascade multilevel inverters with nonequal dc sources using particle swarm optimization,'' \emph{IEEE Transactions on Industrial Electronics}, vol.~57, no.~11, pp. 3678--3684, 2010.

\bibitem{baghaee2017}
H.~R. Baghaee \emph{et~al.}, ``A hybrid anfis/abc-based online selective harmonic elimination switching pattern for cascaded multi-level inverters of microgrids,'' \emph{IEEE Transactions on Industrial Electronics}, 2017.

\bibitem{zolfaghari2024minimizing}
H.~Zolfaghari, H.~Karimi, A.~Ramezani, and M.~Davoodi, ``Minimizing voltage ripple of a dc microgrid via a particle-swarm-optimization-based fuzzy controller,'' \emph{Algorithms}, vol.~17, no.~4, p. 140, 2024.

\bibitem{zolfaghari2022numerical}
A.~Zolfaghari, E.~Aminian, and H.~Saffari, ``Numerical investigation on entropy generation in the dropwise condensation inside an inclined pipe,'' \emph{Heat Transfer}, vol.~51, no.~1, pp. 551--577, 2022.

\bibitem{zolfaghari2021real}
H.~Zolfaghari, D.~Momeni, and H.~Karimi, ``Real time simulation of gird-connected photovoltaic multilevel inverter using hybrid ga/pso optimization algorithm.'' \emph{arXiv preprint arXiv}, vol. 2110, 2021.

\bibitem{zolfaghari2021multilevel}
H.~Zolfaghari, H.~Momeni, and H.~Karimi, ``Multilevel inverter real-time simulation and optimization through hybrid ga/pso algorithm,'' \emph{arXiv preprint arXiv:2110.13817}, 2021.

\bibitem{zolfaghari2021voltage}
H.~Zolfaghari, H.~Karimi, and D.~H. Momeni, ``Voltage stabilization of a dc-microgrid using anfis controller considering electrical vehicles and transient storage,'' \emph{arXiv preprint arXiv:2110.11725}, 2021.

\bibitem{zolfagharivoltage}
H.~Zolfaghari, H.~Karimi, and H.~Momeni, ``Voltage stabilization of a dc-microgrid using anfis controller considering evs, der, and transient storage,'' \emph{arXiv preprint arXiv:2110.11725}, 2021.

\bibitem{turing1950}
A.~M. Turing, ``Computing machinery and intelligence,'' \emph{Mind}, vol.~59, no. 236, pp. 433--460, 1950.

\bibitem{karimi2016}
H.~Karimi and R.~Dashti, ``Comprehensive framework for capacitor placement in distribution networks from the perspective of distribution system management in a restructured environment,'' \emph{International Journal of Electrical Power \& Energy Systems}, vol.~82, pp. 11--18, 2016.

\bibitem{mitchell1998}
M.~Mitchell, \emph{An Introduction to Genetic Algorithms}.\hskip 1em plus 0.5em minus 0.4em\relax MIT Press, 1998.

\bibitem{mirjalili2014}
S.~Mirjalili, G.-G. Wang, and L.~d.~S. Coelho, ``Binary optimization using hybrid particle swarm optimization and gravitational search algorithm,'' \emph{Neural Computing and Applications}, vol.~25, no.~6, pp. 1423--1435, 2014.

\bibitem{yin2006}
P.-Y. Yin, ``Genetic particle swarm optimization for polygonal approximation of digital curves,'' \emph{Pattern Recognition and Image Analysis}, vol.~16, no.~2, pp. 223--233, 2006.

\bibitem{soleimani2015}
H.~Soleimani and G.~Kannan, ``A hybrid particle swarm optimization and genetic algorithm for closed-loop supply chain network design in large-scale networks,'' \emph{Applied Mathematical Modelling}, vol.~39, no.~14, pp. 3990--4012, 2015.

\bibitem{salehi2011}
R.~Salehi \emph{et~al.}, ``Elimination of low order harmonics in multilevel inverters using genetic algorithm,'' \emph{Journal of Power Electronics}, vol.~11, no.~2, pp. 132--139, 2011.

\end{thebibliography}

\end{document}